\documentclass[aps, prl, twocolumn,showkeys, showpacs,amsmath,amssymb,superscriptaddress]{revtex4-1}
\usepackage{graphicx}
\usepackage{dcolumn}
\usepackage{bm}
\usepackage{epsfig}
\usepackage{subfigure}
\usepackage{color}

\begin{document}
\title{Flow Through Randomly Curved Manifolds}

\author{M. Mendoza} \email{mmendoza@ethz.ch} \affiliation{ ETH
  Z\"urich, Computational Physics for Engineering Materials, Institute
  for Building Materials, Schafmattstrasse 6, HIF, CH-8093 Z\"urich
  (Switzerland)}

\author{S. Succi} \email{succi@iac.cnr.it} \affiliation{Istituto per
  le Applicazioni del Calcolo C.N.R., Via dei Taurini, 19 00185, Rome
  (Italy),\\and Freiburg Institute for Advanced Studies,
  Albertstrasse, 19, D-79104, Freiburg, (Germany)}

\author{H. J. Herrmann}\email{hjherrmann@ethz.ch} \affiliation{ ETH
  Z\"urich, Computational Physics for Engineering Materials, Institute
  for Building Materials, Schafmattstrasse 6, HIF, CH-8093 Z\"urich
  (Switzerland)} \affiliation{Departamento de F\'isica, Universidade
  Federal do Cear\'a, Campus do Pici, 60455-760 Fortaleza, Cear\'a,
  (Brazil)}

\date{\today}
\begin{abstract}
  We have found that the relation between the flow through campylotic
  (generically curved) media, consisting of randomly located curvature
  perturbations, and the average Ricci scalar of the system exhibits
  two distinct functional expressions (hysteresis), depending on
  whether the typical spatial extent of the curvature perturbation
  lies above or below the critical value maximizing the overall Ricci
  curvature. Furthermore, the flow through such systems as a function
  of the number of curvature perturbations presents a sublinear
  behavior for large concentrations due to the interference between
  curvature perturbations that, consequently, produces a less curved
  space. For the purpose of this study, we have developed and
  validated a lattice kinetic model capable of describing fluid flow
  in arbitrarily curved manifolds, which allows to deal with highly
  complex spaces in a very compact and efficient way.
\end{abstract}


\maketitle

Many systems in Nature present either intrinsic spatial curvature,
e.g. curved space, due to presence of stars and other interestellar
media \cite{landau}, or geometric confinement constraining the degrees
of freedom of particles moving on such media, e.g. flow on soap films
\cite{soap}, solar photosphere \cite{solar}, flow between two rotating
cylinders and spheres \cite{tay1, transition, taylorsphe1}, to name
but a few.  In general, these systems force a fluid to move along
non-straight trajectories (curved geodesics), leading to the upsurge
of non-inertial forces. We will denote such systems as {\it
  Campylotic}, from the greek word $\kappa \alpha \mu \pi
\acute{\upsilon} \lambda o \varsigma$ for curved, media.  Due to the
arbitrary trajectories that particles through a campylotic medium can
take, depending on the complexity of the curved space, the flow
through these media can present very unusual new transport
properties. Campylotic media play a prominent role in all applications
where metric curvature has a major impact on the flow structure and
topology; biology, astrophysics and cosmology offering perhaps the
most natural examples. Indeed, for several special cases, the flow
through simple campylotic media has already been studied,
e.g. Taylor-Couette flow, which was originally formulated between two
concentric, rotating cylinders \cite{tay1, transition}, and later
extended to the case of spheres \cite{tay3}.  However, beyond very
simple geometries, the flow through more complicated structures, like
randomly located stars or many biological systems, to the best of our
knowledge, has never systematically been addressed before on
quantitative grounds. 
\begin{figure}
  \centering
  \includegraphics[scale=0.185]{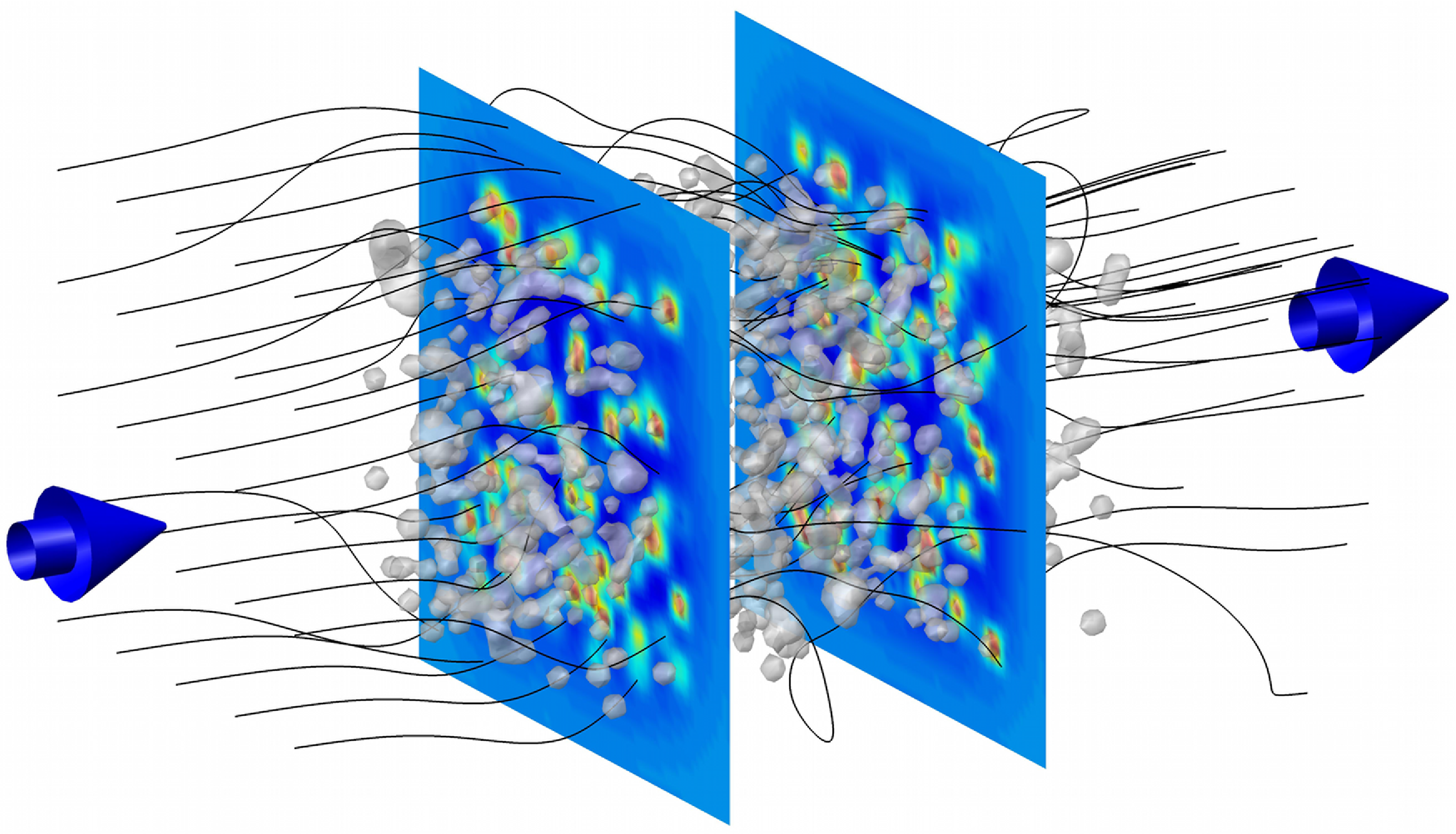}
  \caption{Streamlines of a three-dimensional fluid moving through a
    campylotic medium. The colors denote the Ricci scalar $R'$ (blue
    and red for low and high values, respectively). The gray bubbles
    isosurfaces stand at $1/5$ of the maximum curvature of the
    system.}\label{porous}
\end{figure}
Since, in general, this class of flows lacks analytical solutions,
their study is inherently dependent on the availability of appropriate
numerical methods. Flows in complex geometries, such as cars or
airplanes, make a time-honored mainstream of computational fluid
dynamics (CFD), a discipline which has made tremendous progress for
the last decades \cite{CFD1,CFD2}. However, campylotic media set a
major challenge even to the most sophisticated CFD methods, because
the geometrical complexity is often such to command very high spatial
accuracy to resolve the most acute metric and topological features of
the flow. Therefore, in this work, we also present a new lattice
kinetic scheme that can handle flows in virtually arbitrary complex
manifolds in a very natural and elegant way, by resorting to a
covariant formulation of the lattice Boltzmann (LB) kinetic equation
in general coordinates. The method is validated quantitatively for
very simple campylotic media by calculating the critical Reynolds
number for the onset of the Taylor-Couette instability in concentric
cylinders and spheres \cite{transition, tay3,taylorsphe1,taylorsphe2},
and applied to the case of two concentric tori.

In this Letter, by using the new numerical scheme, we simulate the
flow through campylotic media consisting of randomly distributed
spatial curvature perturbations (see Fig.~\ref{porous}). The flow is
characterized by the number of curvature perturbations and the average
Ricci scalar of the space. The campylotic media explored in this work
are static, in the sense that the metric tensor and curvature are
prescribed at the outset once and for all, and do not evolve
self-consistently with the flow. The latter case, which is a major
mainstream of current numerical relativity \cite{numRel1, numRel2},
makes a very interesting subject for future extensions of this work.

In order to study the campylotic media, we develop a lattice kinetic
approach in general geometries, taking into account the metric tensor
$g_{ij}$ and the Christoffel symbols $\Gamma_{k j}^i$. The former
characterizes the way to measure distances in space, while the latter
is responsible for the non-inertial forces. The corresponding
hydrodynamic equations can be obtained by replacing the partial
derivatives by covariant ones, in both, the mass continuity and the
momentum conservation equations. After some algebraic manipulations,
the hydrodynamic equations read as follows: $\partial_t \rho + (\rho
u^i)_{;i} = 0$, and $\partial_t (\rho u^i) + T^{ij}_{;j} = 0$, where
the notation $_{;i}$ denotes the covariant derivative with respect to
spatial component $i$ (further details are given in the Supplementary
Material \cite{supp}).  The energy tensor $T^{ij}$ is given by,
$T^{ij} = P g^{ij} + \rho u^i u^j - \mu (g^{lj} u^i_{;l} + g^{il}
u^j_{;l} + g^{ij} u^l_{;l})$, where $P$ is the hydrostatic pressure,
$u^i$ the $i$-th contravariant component of the velocity, $g^{ij}$ the
inverse of the metric tensor, $\rho$ is the density of the fluid, and
$\mu$ is the dynamic shear viscosity.

Since lattice Boltzmann methods are based on kinetic theory, we
construct our model by writing the Maxwell-Boltzmann distribution and
the Boltzmann equation in general geometries.  The former takes the
form \cite{donato}:
\begin{equation}\label{MBdistribution}
  f^{\rm eq} = \frac{\sqrt{g} \rho}{\left( 2\pi \theta\right)^{3/2}} \exp \left[ -\frac{1}{2\theta} g_{ij} (\xi^i - u^i)(\xi^j - u^j) \right] \quad ,
\end{equation}
where $g$ is the determinant of the metric $g_{ij}$, and $\theta$ is
the normalized temperature. The macroscopic and microscopic
velocities, $u^i$ and $\xi^i$ are both normalized with the speed of
sound $c_s = \sqrt{k_B T_0/m}$, $k_B$ being the Boltzmann constant,
$T_0$ the typical temperature, and $m$ the mass of the particles. Note
that the metric tensor appears explicitly in the distribution
function, due to the fact that the kinetic energy is a quadratic
function of the velocity, $u^i u_i = g_{ij}u^i u^j$.  To recover the
macroscopic fluid dynamic equations, we have to extract the moments
from the equilibrium distribution function. The four first moments of
the Maxwellian distribution function on a manifold are given by,
\begin{subequations}\label{moments}
\begin{equation}
  \rho = \int f  d\xi \quad , \quad \rho u^i = \int f \xi^i d\xi \quad ,
\end{equation}
\begin{equation}
  \rho \theta g^{ij} + \rho u^i u^j = \int f \xi^i \xi^j d\xi \quad ,
\end{equation}
\begin{equation}\label{moments:3}
  \rho \theta ( u^i g^{jk} + u^j g^{ik} + u^k g^{ij} ) + \rho u^i u^j
  u^k = \int f \xi^i \xi^j \xi^k d\xi  .
\end{equation}
\end{subequations}
These moments are sufficient to reproduce the mass and the momentum
conservation equations. Here, for simplicity we have used $d\xi$ to
denote $d\xi^1 d\xi^2 d\xi^3$ and the Jacobian of the integration is
already included in the Maxwell Boltzmann distribution, through the
determinant term $\sqrt{g}$.

In the absence of external forces, in the standard theory of the
Boltzmann equation, the single particle distribution function $f(x^i,
\xi^i, t)$ evolves, according to the equation, $\partial_t f + \xi^i
\partial_i f = {\cal C}(f)$, where $\cal C$ is the collision term,
which, using the BGK approximation, can be written as, ${\cal C} =
-(1/\tau) (f - f^{\rm eq})$, with the single relaxation time $\tau$.
This equation can be obtained from a more general expression, $df/dt =
{\cal C}(f)$, where the total time derivative now includes a streaming
term in velocity space due to external forces, $\frac{df}{dt} =
\partial_tf + \frac{dx^i}{dt}\partial_i f + \frac{dp^i}{dt}
\partial_{p^i} f$, with $p^i$ the $i$-th contravariant component of
the momentum of the particles. Using the definition of velocity,
$\xi^i = dx^i/dt$, and due to the fact that the particles in our fluid
move along geodesics, which implies the equation of motion 
\begin{equation}\label{eqgeo}
  \frac{dp^i}{dt} =-\Gamma^i_{kl} p^k p^l \quad ,
\end{equation}
we can write the Boltzmann equation as
\cite{KineticBoltzmann},
\begin{equation}\label{Boltzmann3:eq}
  \partial_t f + \xi^i \partial_i f - \Gamma^{i}_{jk} \xi^j
  \xi^k \partial_{\xi^i} f = {\cal C}(f) \quad ,
\end{equation}
where we have used the definition of the momentum, $p^i = m \xi^i$.
Note that the third term of the left hand side carries all the
information on non-inertial forces. Thus, all the ingredients required
to model a fluid in general geometries within the Boltzmann equation
are now in place. Note that the Christoffel symbols and metric tensor
are arbitrary and therefore we can model the fluid flow in curved
spaces, whose metric tensor is very complicated and/or only known
numerically.

Since the contravariant components of the velocity are free of
space-dependent metric factors, they lend themselves to standard
lattice Boltzmann discretization of velocity space. All the metric and
non-inertial information is conveyed into the generalized local
equilibria and forcing term, respectively. These features are key to
the LB formulation in general manifolds. As an additional feature,
complex boundary conditions related to a specific geometry, e.g.
surface of sphere, in many cases, can be treated exactly by cubic
cells in the contravariant coordinate frame, thereby avoiding
stair-case approximations typical of cartesian grids. The
details of the discretization of this model on a lattice can be found
in the Supplementary Material \cite{supp}.

To provide numerical validation of our model, we study the flow
through one of the simplest campylotic medium, the Taylor-Coutte
instability in three different geometries, i.e.  two concentric
rotating cylinders, spheres and tori, respectively.  Full details of
the validation are given in the Supplementary Material\cite{supp}.  In
Figure \ref{taylor:fig1}, we report the critical Reynolds number as a
function of the aspect ratio $\eta=a/b$, where $a$ and $b$ are the
minor and major radii, respectively.  As one can appreciate, for the
cylindrical geometry we obtain excellent agreement with analytical
theory \cite{transition}, and a similar match with experimental data
\cite{tay3} is found for the spherical case. We have also computed the
torque coefficient, and found reasonable agreement, within a few
percent, with experimental data \cite{torque1, torque2}.  For the case
of two concentric rotating tori, the critical Reynolds numbers for
different configurations can also be observed in
Fig.~\ref{taylor:fig1}, showing values around $10\%$ larger than for
the case of cylinders. Further details can be found in the
Supplementary Material \cite{supp}.

\begin{figure}
  \centering
  \includegraphics[scale=0.33]{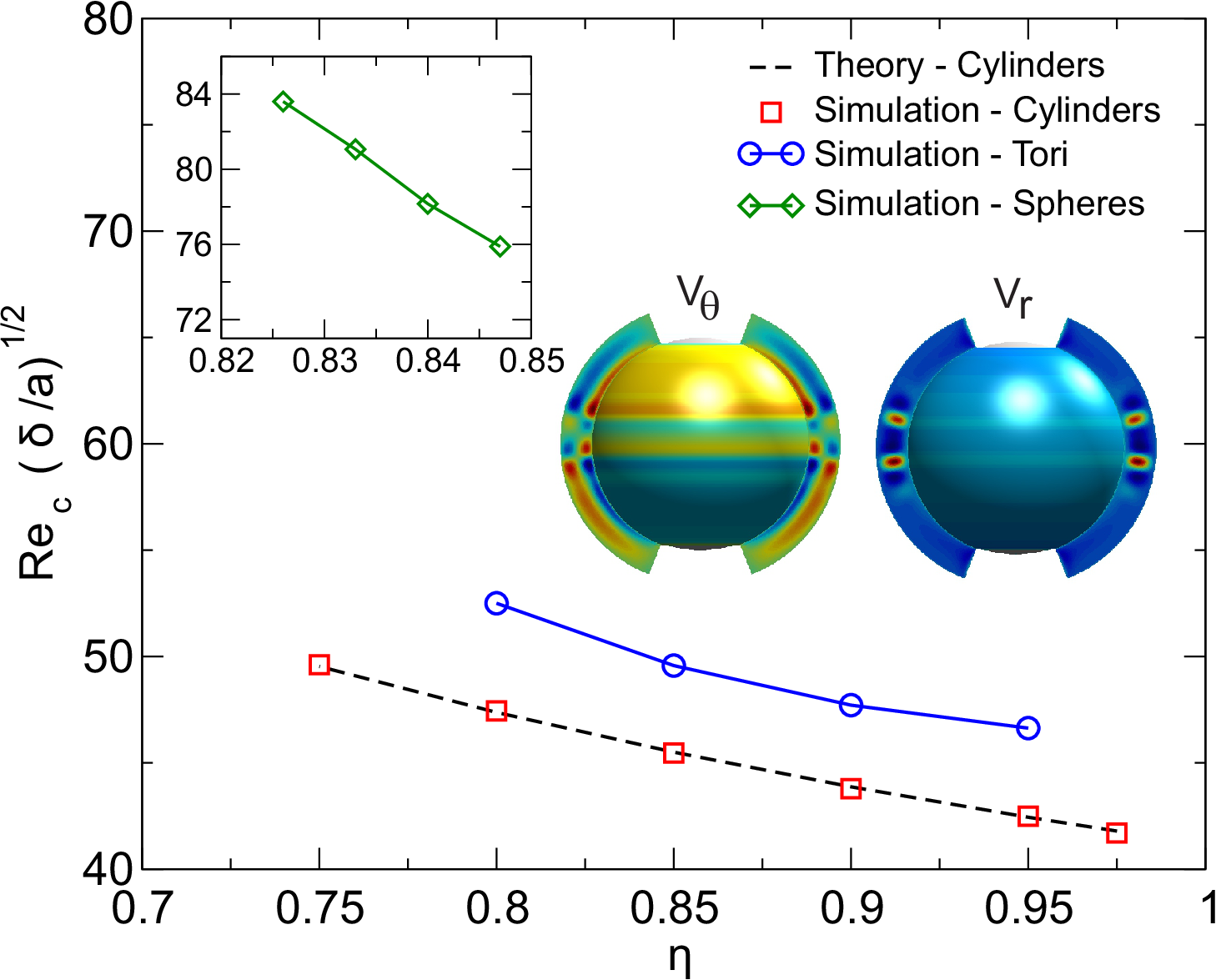}
  \caption{Critical Reynolds number $Re_c$, as a function of the
    parameter $\eta=a/b$ at the onset of the Taylor-Couette
    instability, for two concentric rotating cylinders (red) and tori
    (blue). Theoretical values for the case of the cylinders agree
    with Ref.~\cite{transition}. The left inset shows the critical
    Reynolds number for the case of two concentric spheres, and the
    two colored spheres the radial and axial components of the fluid
    velocity for the spherical case. Blue and red colors denote low
    and high values, respectively.}
  \label{taylor:fig1}
\end{figure}
\begin{figure}
  \centering
  \includegraphics[scale=0.38]{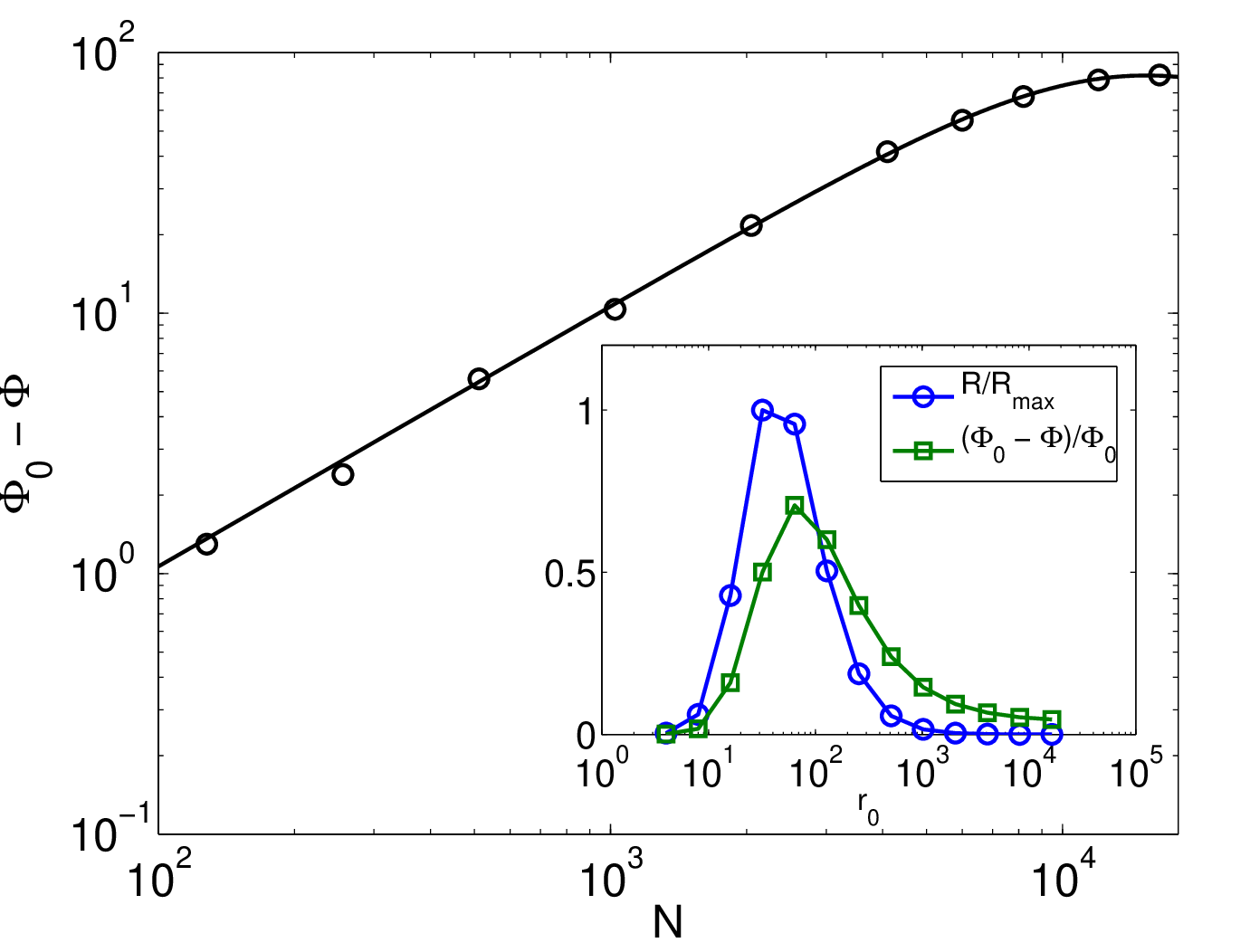}
  \caption{Flux reduction $\Phi_0-\Phi$ with respect to the flat case,
    as a function of the number of curvature perturbations for $a_0 =
    0.01$ and $r_0 = 2.0$. The solid line is the analytical curve
    according to Eq.~\eqref{camp:eq0}. Shown in the inset is the
    normalized average curvature scalar of the space, $R/R_{max}$, and
    the normalized reduced flux $1-\Phi_0/\Phi$ as a function of
    $r_0$.  Both Ricci scalar and flux reduction exhibit a maximum at
    intermediate values of $r_0$. Since the two maxima are slightly
    shifted with respect to each other, the reduced flow as a function
    of $R$ exhibits an hysteresis loop (see next Figure
    \ref{camp:fig2}).  }
  \label{camp:fig1}
\end{figure}
\begin{figure}
  \centering
  \includegraphics[scale=0.38]{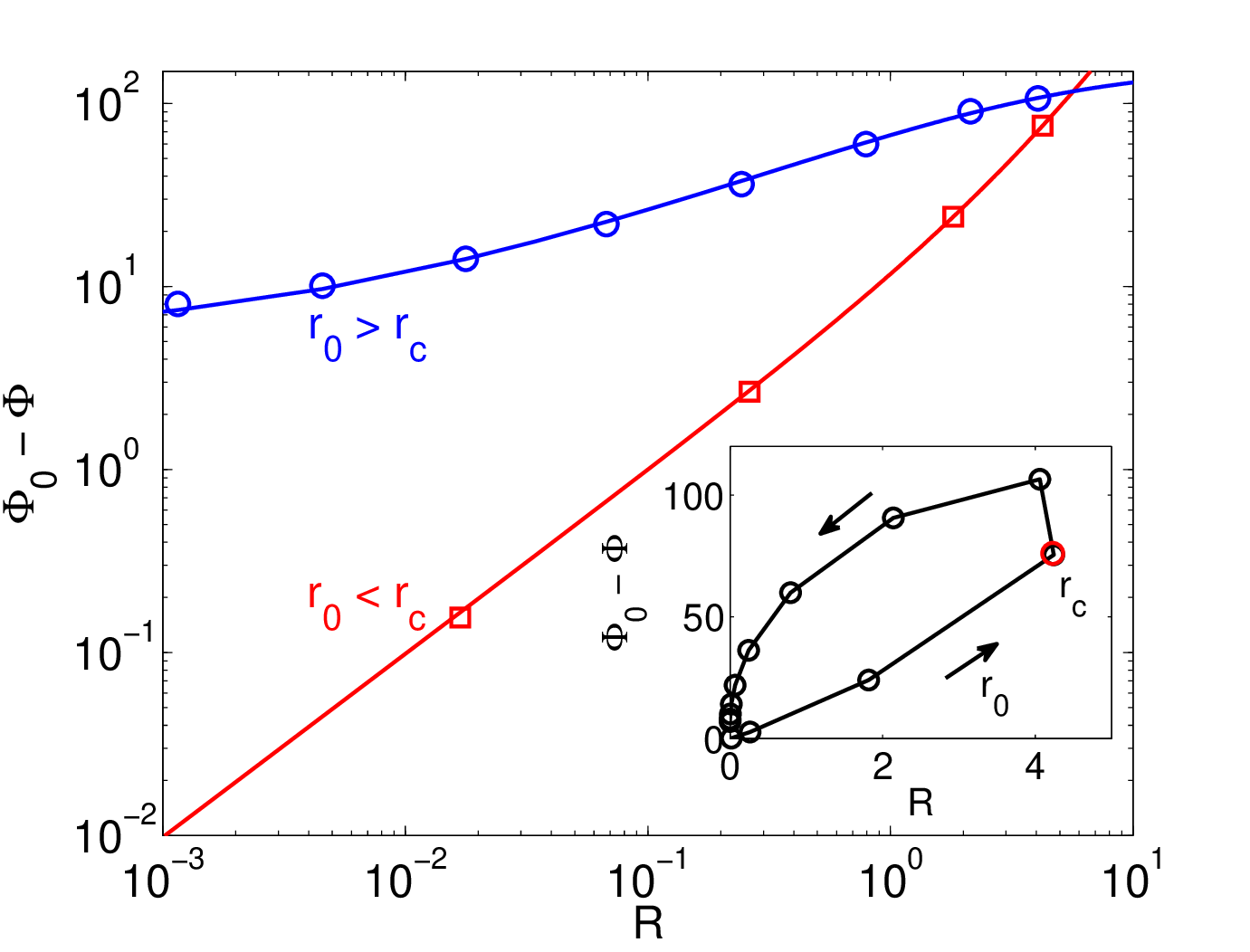}
  \caption{Flux reduction, $\Phi_0-\Phi$, as a function of the average
    curvature, $R$, for large and small values of $r_0$.  We have
    fixed $a_0 = 0.00002$ and $N = 1024$.  The solid lines denote the
    analytical curves according to Eqs.~\eqref{camp:eq1} and
    ~\eqref{camp:eq2}.  The inset shows the hysteresis loop which
    arises by parametrizing the flux-curvature relation in terms of
    $r_0$.  Here, $r_c$ is the radius at which the Ricci curvature
    attains its maximum upon increasing $r_0$.  The lower and upper
    branches correspond to $r_c<r_0$ and $r_c>r_0$, respectively.  }
  \label{camp:fig2}
\end{figure}

Next, we move to a genuinely campylotic medium, consisting of randomly
located curvature perturbations.  To this purpose, we define a
coordinate system $(x,y,z)$, such that its metric tensor takes the
form: $g_{ij} = \delta_{ij}( 1 - a_0\sum_{n=1}^N \exp(-r_n/r_0))$,
where $n$ labels each local curvature perturbation located at
$\vec{r}_n = (x_n,y_n,z_n)$, $N$ is the total number of perturbations,
$r_n = |\vec{r}_n|$, and $r_0$ characterizes the size of the
deformation. Note that the coefficient $a_0$ can be either signed,
depending on whether a positive or negative curvature is imposed,
respectively.  In our study, we have chosen to work with positive
values of $a_0$, due to the analogy with a system of randomly located
stars, which produce deformations in the metric tensor of
spacetime\cite{landau}. The Christoffel symbols are calculated
numerically. The flux is calculated by the geometrical relation, $\Phi
= \int_S \rho u^x \sqrt{g^{xx} g } dS$, where $S$ is the cross section
at the location where the measurements are taken.  Since the fluid
dynamic equations only contain the metric tensor and its first
derivatives (via the Christoffel symbol), and due to the fact that
particles move along geodesics according to Eq.~\eqref{eqgeo}, it is
natural to expect that the flow could be characterized by a quantity
that contains the metric tensor and its first derivatives.  Although
the Christoffel symbols $\Gamma^i_{jk}$ meet this requirement, they
are not components of a tensor, and therefore they are not invariant
under a coordinate system transformation (physics should not depend on
the choice of the coordinate system).  An invariant, or tensor, that
can be used to characterize the system is the Ricci tensor
$R_{ik}$. In this work, we use the Ricci scalar (curvature scalar)
which can be calculated from the Ricci tensor, $R_{ij}$, by
contraction of the indices, $R' = g^{ij} R_{ij}$. The relation between
the metric tensor and Christoffel symbols and the Ricci tensor can be
found in the Supplementary Material \cite{supp}.  To study this
particular system, we use a lattice size $L_x\times L_y \times L_z$ of
$128\times 64\times 64$, and $\tau = 1$. All quantities will be
expressed in numerical units.  To drive the fluid through the medium,
we add an external force along the $x$-component, which in all
simulations takes the value, $f_{ext} = 5\times 10^{-5}$. The flux in
flat space, i.e. in the absence of curvature perturbations is denoted
by $\Phi_0$.

Shown in Fig.~\ref{porous}, are the velocity streamlines, the Ricci
scalar $R'$ and the high-curvature locations, represented by gray
isosurfaces. Note that the streamlines are very complex, as the flow
can orbit around the spheres before continuing its trajectory
\cite{BINI1,BINI2}. Also we can see how the curvature perturbations
interact, creating non-spherical shaped isosurfaces.

Fig.~\ref{camp:fig1} shows the flux reduction $\Phi_0-\Phi$, as
function of the number of curvature perturbations, $N$. We observe
that the flux $\Phi$ decreases with $N$.  This effect is due to the
interplay between the longer trajectories that particles must take and
the acceleration due to the non-inertial forces, see
Eq.~\eqref{eqgeo}. Note that, in general, for systems with different
configurations (e.g. negative $a_0$), we could expect that the
combination of the two effects might lead to higher flux by increasing
$N$. We also see that the flux depends linearly on $N$ for low
concentration of curvature perturbations, and only sublinearly at
higher concentrations.  This is due to the fact that at low
concentration, the average distance between curvature perturbations is
large, and consequently each perturbation adds up as a single
modification to the total spatial curvature.  However, as the
concentration is increased, the curvature perturbations start to
interfere with each other and consequently the space becomes less
curved (decrease of the overall Ricci curvature).  The flux is found
to obey the following law,
\begin{equation}\label{camp:eq0}
  \Phi_0 - \Phi = A_1 \frac{N/N_0}{1+(N/N_0)^2} \quad , 
\end{equation}
where $A_1 = 163 \pm 2$ and $N_0 = (1.54 \pm 0.03)\times 10^4$ are
fitting parameters. The parameter $N_0$ denotes a characteristic
number of curvature perturbations, above which the sublinear behavior
sets in ($N \gtrsim N_0$).

In the inset of Fig.~\ref{camp:fig1}, we observe that by fixing the
number of curvature perturbations $N = 1024$ and the strength $a_0 =
0.01$, and changing the range of the perturbation, $r_0$, the
difference $\Phi_0 - \Phi$ presents a maximum for a given $r_0 \sim
r_c$. Furthermore, another interesting result is that the average
curvature, here defined as $R = - 10^8 <R'>$ (where $<...>$ means
average over space), shows the same qualitative behavior. Since by
increasing $r_0$ the metric tensor components decrease monotonically,
this maximum is due to the Christoffel symbols (or non-inertial
forces), which can be characterized via $R$. However, the maxima are
slightly shifted, due to the fact that the Ricci scalar does not
uniquely determine the metric tensor and Christoffel symbols, the
quantities that play a key role in the fluid dynamic equations. Taking
into account this effect, we can plot the flux reduction $\Phi_0 -
\Phi$ as a function of $R$, and find that, indeed, for $r_0 < r_c$,
the flux decreases by increasing the average curvature $R$ with a
different law than for the case of large values of $r_0 > r_c$ (see
inset of Fig.~\ref{camp:fig2}).  This gives rise to a
hysteresis-shaped curve, the reason for this hysteresis being that the
metric tensor is different for $r_0 < r_c$ and $r_0 > r_c$, even if
$R$ takes the same value. However, in both cases, the system shows
that higher values of the average curvature $R$ always result in a
lower flux.  The behavior of the flux for $r_0 < r_c$ is well
represented by the following law:
\begin{equation}\label{camp:eq1}
  \Phi_0 - \Phi = A_2 \frac{R}{R_0}\left ( 1 + \frac{R}{R_0}\right ) \quad ,
\end{equation}
and for the case of $r_0 > r_c$,
\begin{equation}\label{camp:eq2}
  \Phi_0 - \Phi = A_3 \sqrt{\frac{R}{R_0 + R}} + \Phi' \quad ,
\end{equation}
where $R_0 = 5.2 \pm 0.1$, $A_2 = 50 \pm 2$, $A_3 = 154 \pm 4$, and
$\Phi' = 5 \pm 1$.  The quantity $R_0$ is related to the maximum
curvature achieved by the system and the intersection of the two laws
(see Fig.~\ref{camp:fig2}).  The other interesting quantity is
$\Phi'$, which represents the difference of flux between $r_0 \gg r_c$
and $r_0 \ll r_c$, when the curvature scalar becomes zero, and it is
due to the fact that in both cases, although the space has no
curvature, it has nonetheless different metric tensors.

Summarizing, we have explored the laws that rule the flow through
campylotic media consisting of randomly distributed curvature
perturbations, and shown that, for the configurations studied in this
Letter, curved spaces invariably support less flux than flat spaces.
Furthermore, the flux can be characterized by the Ricci scalar, a
geometrical invariant that contains the metric tensor and Christoffel
symbols, the quantities that appear in the fluid dynamics equations.
The trajectories of the flow can become very complicated due to the
total curvature of the medium, presenting, in some cases, orbits
winding several times around regions with high curvature.  The present
method opens the possibility to apply the actual model to
astrophysical systems, where the curvature of space is due to the
presence of stars and other interstellar material.  We have not
considered time curvature, since its contribution remains sub-dominant
unless mass is made extremely large.

To calculate the flux in campylotic media, we have developed a new
lattice Boltzmann model to simulate fluid dynamics in general
non-cartesian manifolds. The model has been successfully validated on
the Taylor-Couette instability for the case of two concentric
cylinders and spheres, the inner rotating with a given speed and the
outer being fixed. We also studied the Taylor-Couette instability in
two concentric rotating tori, finding that the critical Reynolds
number for the onset of the instability is about ten percent larger
than the one for the cylinder. By solving the Navier-Stokes equations
in contravariant coordinates, which can be represented on a cubic
lattice precisely in the format requested by the lattice Boltzmann
formulation, the present model opens up the possibility to study fluid
dynamics in complex manifolds by retaining the outstanding simplicity
and computational efficiency of the standard lattice Boltzmann method
in cartesian coordinates.  The case of dynamically adaptive campylotic
media, in which the metric tensor and curvature would evolve
self-consistently together with the flow, makes a very interesting
subject for future extensions of the present lattice kinetic method in
the direction of numerical relativity \cite{rlbPRL, turbPRL}.

\begin{acknowledgments}
  The authors are grateful for the financial support of the
  Eidgen\"ossische Technische Hochschule Z\"urich (ETHZ) under Grant
  No. 06 11-1.
\end{acknowledgments}

\appendix
\section{Supplementary Material}

We show the details of the new lattice kinetic model to study
campylotic media, and include a respective validation by studying the
Taylor-Couette instability for the case of two concentric rotating
cylinders, spheres and tori. We also implement a convergence study
showing that the model presents nearly second order convergence, and
introduce basic relations in differential geometry like the
calculation of covariant derivatives and the Ricci tensor.

\section{Covariant derivative and Ricci Tensor}

The formulation of fluid equations in general coordinates implies the
replacement of partial derivatives with the corresponding covariant
ones.  Given a vector $A^i$, the covariant derivative is defined by
\begin{equation}
  A^i_{;j} = \partial_j A^i + \Gamma^i_{jk}A^k \quad ,
\end{equation}
where $\Gamma^i_{jk}$ is the Christoffel symbol associated with the
curvature of the metric manifold, namely $\Gamma^i_{jk} = \frac{1}{2}
g^{im} ( \frac{\partial g_{jm}}{\partial x^k}+ \frac{\partial
  g_{km}}{\partial x^j} -\frac{\partial g_{jk}}{\partial x^m}) $.  For
an arbitrary tensor of second order, the covariant derivative is given
by
\begin{equation}
 A^{ik}_{;l} = \partial_l A^{ik} + \Gamma^i_{ml}A^{mk} + \Gamma^k_{ml}A^{im} \quad .
\end{equation}
Here and throughout, according to Einstein's convention, repeated
indices are summed upon. 

The Ricci tensor $R_{ik}$ is related with the metric tensor and
Christoffel symbols by the relation,
\begin{equation}
  R_{ik} = \frac{\partial \Gamma_{ik}^l}{\partial x^l} - \frac{\partial \Gamma_{il}^l}{\partial x^k} + \Gamma_{ik}^l\Gamma_{lm}^m - \Gamma_{il}^m \Gamma_{km}^l \quad .
\end{equation}

\section{Tensor Hermite Polynomials}

The Lattice Boltzmann formulation in general geometries makes strong
reliance on Hermite expansion of the kinetic distribution function.
The first three Hermite polynomials are,
\begin{subequations}\label{hermite}
  \begin{equation}
    H_{(0)} = 1 \quad ,
  \end{equation}
  \begin{equation}
    H_{(1)}^i = \xi^i \quad ,
  \end{equation}
  \begin{equation}
    H_{(2)}^{ij} = \xi^i \xi^j - \delta^{ij} \quad ,
  \end{equation}
  \begin{equation}
    H_{(3)}^{ijk} = \xi^i \xi^j \xi^k - (\delta^{ij} \xi^k +
    \delta^{kj} \xi^i + \delta^{ik} \xi^j ) \quad ,
  \end{equation}  
\end{subequations}
where we have used the Kronecker delta $\delta^{ij}$. 

\section{Hermite polynomials expansion}

Let us expand the distribution function $f(x^i, \xi^i,t)$ in the form,
\begin{equation}\label{expansion0}
  f(x^i, \xi^i, t) = w(\xi) \sum_{n=0}^{\infty} \frac{1}{n!} {a_{(n)}}(x^i,t){H_{(n)}}(\xi^i) \quad ,
\end{equation}
where the coefficients $a_{(n)}$ are $n$-th order space-time dependent
tensors, and $H_{(n)}$ are the tensorial Hermite polynomials of $n$-th
order.  The weights $w(\xi)$ are defined as:
\begin{equation}
  w(\xi) = \frac{1}{(2\pi)^{3/2}} \exp (-\xi^2/2)\quad .
\end{equation}
The coefficients $a_{(n)}$ can be calculated with the relation,
\begin{equation}\label{coeff}
  a_{(n)} = \int f H_{(n)}(\xi) d\xi \quad .
\end{equation}
To recover the correct hydrodynamic equations, the model must be built
in such a way as to recover the first four moments, given by,
\begin{subequations}\label{moments}
\begin{equation}
  \rho = \int f  d\xi \quad , \quad \rho u^i = \int f \xi^i d\xi \quad ,
\end{equation}
\begin{equation}
  \rho \theta g^{ij} + \rho u^i u^j = \int f \xi^i \xi^j d\xi \quad ,
\end{equation}
\begin{equation}\label{moments:3}
  \rho \theta ( u^i g^{jk} + u^j g^{ik} + u^k g^{ij} ) + \rho u^i u^j
  u^k = \int f \xi^i \xi^j \xi^k d\xi  .
\end{equation}
\end{subequations}

The fourth one ensures that the dissipation term achieves the correct
form. To this purpose, we need to expand the distribution function at
least up to the third order Hermite polynomial (The explicit
expression of the Hermite polynomials have been given above).  Thus,
using Eq.~\eqref{coeff}, and replacing the Maxwell-Boltzmann
distribution for a manifold, we obtain:
\begin{equation}\label{MBdistribution}
  f^{\rm eq} = \frac{\sqrt{g} \rho}{\left( 2\pi \theta\right)^{3/2}} \exp \left[ -\frac{1}{2\theta} g_{ij} (\xi^i - u^i)(\xi^j - u^j) \right] \quad ,
\end{equation}
Next, by taking $\theta = 1$ (isothermal limit), we obtain the
coefficients of the expansion, as follows:
\begin{subequations}\label{coeffh}
  \begin{equation}
    a_{(0)} = \rho \quad ,   \quad a_{(1)}^i = \rho u^i \quad ,
  \end{equation}
  \begin{equation}
    a_{(2)}^{ij} = g^{ij} - \delta^{ij} + \rho u^i u^j \quad ,
  \end{equation}
  \begin{equation}
    a_{(3)}^{ijk} =  (g^{ij}-\delta^{ij}) u^k +
    (g^{kj}-\delta^{kj}) u^i + (g^{ik}-\delta^{ik}) u^j  + \rho u^i u^j u^k .
  \end{equation}  
\end{subequations}
Therefore, the truncated equilibrium distribution function up to third
order, using Eq.~\eqref{expansion0}, reads as follows:
\begin{widetext}
  \begin{equation}
    \begin{aligned}
      f^{\rm eq} = w(\xi) \rho \biggl( \frac{5}{2} &+ 2 \xi^i u^i +
        \frac{1}{2} \xi^i g^{ij} \xi^j - \frac{1}{2}\xi^i \xi^i +
        \frac{1}{2} (\xi^i u^i)^2 - \frac{1}{2} g^{ii} - \frac{1}{2}
        u^i u^i + \frac{1}{6} (\xi^i u^i)^3 \\ & - \frac{1}{2} (\xi^i u^i) 
        (u^j u^j) + \frac{1}{2} (\xi^i u^i)( \xi^j g^{jk} \xi^k -
        \xi^j \xi^j) - \frac{1}{2}(\xi^i u^i)(g^{jj} - 3) - u^i g^{ij}
        \xi^j \biggr) \quad .
    \end{aligned}
  \end{equation}
\end{widetext}
With this, we have expanded the equilibrium distribution
function up to third order in Hermite polynomials.  Next, we need to
expand also the forcing term, $\Gamma^{i}_{jk} \xi^j \xi^k
\partial_{\xi^i} f$, in the Boltzmann equation,
\begin{equation}\label{Boltzmann3:eq}
  \partial_t f + \xi^i \partial_i f - \Gamma^{i}_{jk} \xi^j
  \xi^k \partial_{\xi^i} f = {\cal C}(f) \quad ,
\end{equation}
Due to the fact that the distribution function can be written using
Eq.~\eqref{expansion0}, and invoking the properties of the Hermite
polynomials,
\begin{equation}
  w H_{(n)}^i = (-1)^n \partial_{\xi^i} w \quad ,
\end{equation}
we can write the forcing term as,
\begin{equation}
  F^i \partial_{\xi^i}f =  w \sum_{n=1}^{\infty} \frac{a_{(n-1)}
    F^i}{(n-1)!} H_{(n)}^i \quad ,
\end{equation}
where we have introduced the notation, $F^i = -\Gamma^{i}_{jk} \xi^j
\xi^k$. Then, replacing the coefficients from Eq.~\eqref{coeffh}, and
the corresponding Hermite polynomials, Eq.~\eqref{hermite}, we obtain
the forcing term,
\begin{widetext}
  \begin{equation}
    \begin{aligned}
      -\Gamma^{i}_{jk} \xi^j \xi^k \partial_{\xi^i} f = w(\xi) \rho
      \biggl( \xi^i \xi^j \xi^k \Gamma^i_{jk} &+ (\xi^l u^l)\xi^i
      \xi^j \xi^k \Gamma^i_{jk} - u^i \xi^j \xi^k \Gamma^i_{jk} \\ &+
      \frac{1}{2}(g^{kl} - \delta^{kl} + u^k u^l)( \xi^k \xi^l \xi^i
      \xi^j \xi^m \Gamma^i_{jm} - \xi^k \xi^j \xi^i \Gamma^l_{ji} -
      \xi^l \xi^j \xi^i \Gamma^k_{ji} - \xi^m \xi^j \xi^i
      \Gamma^m_{ji} \delta^{kl}) \biggr) \quad .
    \end{aligned}
  \end{equation}
\end{widetext}

With every term expressed as a series of Hermite polynomials, all is
in place to proceed with the LB discretization according to standard
Hermite-Gauss projection of the continuum Boltzmann equation.

\section{Lattice Discretization}

In order to formulate a corresponding lattice Boltzmann model, we
implement an expansion of the Maxwell-Boltzmann distribution in
Hermite polynomials, so as to recover the moments of the distribution
function up to third order in velocities, as it is needed to correctly
reproduce the dissipation term in the hydrodynamic equations. The
expansion of the Maxwell-Boltzmann distribution was introduced by Grad
in his $13$ moment system \cite{grad}. Since this expansion is
performed in velocity space, and the metric only depends on the
spatial coordinates, we expect such an expansion to preserve its
validity also in the case of a general manifold.  We have followed a
similar procedure as the one described in Refs.~\cite{discLB1,
  discLB2}.

For the discretization of the Maxwell Boltzmann distribution
\eqref{MBdistribution} and the Boltzmann equation
\eqref{Boltzmann3:eq}, we need a discrete velocity configuration
supporting the expansion up to third order in Hermite polynomials.
Our scheme is based on the $D3Q41$ lattice proposed in Ref.~
\cite{karli}, which corresponds to the minimum configuration
supporting third-order isotropy in three spatial dimensions, along
with a H-theorem for future entropic extensions \cite{ELB0} of the
present work.

In the following, we shall use the notation $c_\lambda^i$ to denote
the $i$-th contravariant component of the vector numbered $\lambda$.
Thus, the discrete Boltzmann equation for our model takes the form,
$f_\lambda(x^i + c^i_\lambda \delta t, t + \delta t) - f_\lambda(x^i,
t) =-\frac{\delta t}{\tau} (f_\lambda - f_\lambda^{\rm eq}) + \delta t
{\cal F}_\lambda$, where ${\cal F}_\lambda$ is the forcing term, which
contains the Christoffel symbols, and $f_\lambda^{\rm eq}$ is the
discrete form of the Maxwell-Boltzmann distribution,
Eq.~\eqref{MBdistribution}. The relevant physical information about
the fluid and the geometry of the system is contained in these two
terms. The macroscopic variables are obtained according to the
relations, $\rho = \sum_{\lambda = 0}^{41} f_\lambda$, $\rho u^i =
\sum_{\lambda = 0}^{41} f_\lambda c_\lambda^i$. The shear viscosity of
the fluid can also be calculated as $\mu = \rho (\tau - 1/2) c_s^2
\delta t$.

In the following, we shall use the notation $c_\lambda^i$ to denote
the vector number $\lambda$ and the contravariant component $i$. The
cell configuration $D3Q41$ has the discrete velocity vectors:
$(0,0,0)$, $(\pm 1, 0,0)$, $(\pm 1, \pm 1, 0)$, $(\pm 1, \pm 1, \pm
1)$, $(\pm 3, 0, 0)$, $(0, \pm 3, 0)$, $(0, 0, \pm 3)$, and $(\pm 3,
\pm 3, \pm 3)$. The speed of sound for this configuration is $c_s^2 =
1 - \sqrt{2/5}$.  With this setup, and taking into account that the
vectors $\xi^i$ and $u^i$ are normalized by the speed of sound, we
obtain the following equilibrium distribution,
\begin{equation}\label{feq}
  \begin{aligned}
    f_\lambda^{\rm eq} &= w_\lambda \rho \biggl( \frac{5}{2} + 2
    \frac{c_{\lambda}^i u^i}{c_s^2} + \frac{1}{2} \frac{c_{\lambda}^i
      g^{ij} c_{\lambda}^j}{c_s^2} - \frac{1}{2} \frac{c_{\lambda}^i
      c_{\lambda}^i}{c_s^2} \\ &+ \frac{1}{2} \frac{(c_{\lambda}^i
      u^i)^2}{c_s^4} - \frac{1}{2} g^{ii} - \frac{1}{2} \frac{u^i
      u^i}{c_s^2} + \frac{1}{6} \frac{(c_{\lambda}^i u^i)^3}{c_s^6} \\
    &- \frac{1}{2} \frac{(c_{\lambda}^i u^i) (u^j u^j)}{c_s^4} +
    \frac{1}{2} \frac{(c_{\lambda}^i u^i)}{c_s^4}( c_{\lambda}^j
    g^{jk} c_{\lambda}^k - c_{\lambda}^j c_{\lambda}^j) \\ &-
    \frac{1}{2}\frac{(c_{\lambda}^i u^i)}{c_s^2}(g^{jj} - 3) -
    \frac{u^i g^{ij} c_{\lambda}^j}{c_s^2} \biggr) \quad ,
  \end{aligned}
\end{equation}
where the weights $w_\lambda$ are defined as, $w_{(0,0,0)} =
\frac{2}{2025} (5045 - 1507\sqrt{10})$, $w_{(1,0,0)} =
\frac{37}{5\sqrt{10}} - \frac{91}{40}$, $w_{(1,1,0)} = \frac{1}{50}(55
- 17\sqrt{10})$, $w_{(1,1,1)} = \frac{1}{1600}(233\sqrt{10} - 730)$,
$w_{(3,0,0)} = \frac{1}{16200}(295 - 92\sqrt{10})$, and $w_{(3,3,3)} =
\frac{1}{129600}(130 - 41\sqrt{10})$.

The discrete Boltzmann equation for our model takes the form,
\begin{equation}\label{dBoltzmann:eq}
  f_\lambda(x^i + c^i_\lambda \delta t, t + \delta t) - f_\lambda(x^i, t)
  =-\frac{\delta t}{\tau} (f_\lambda - f_\lambda^{\rm eq}) + \delta t {\cal F}_\lambda \quad ,
\end{equation}
where we have introduced the forcing term,
\begin{equation}
  \begin{aligned}
    \delta t {\cal F}_\lambda &= w_\lambda \rho \Biggl[
    \frac{c_{\lambda}^i F_\lambda^i}{c_s^2} + \frac{(c_{\lambda}^l
      u^l)c_{\lambda}^i F_\lambda^i}{c_s^4} - \frac{u^i F_\lambda^i}{c_s^2} \\
    &+ \frac{1}{2} \left(g^{kl} - \delta^{kl} + \frac{u^k
        u^l}{c_s^2}\right) \Biggl( \frac{c_{\lambda}^k c_{\lambda}^l
      c_{\lambda}^i F_\lambda^i}{c_s^4} \\ &- \frac{c_{\lambda}^k
      F_\lambda^l}{c_s^2} - \frac{c_{\lambda}^l F_\lambda^k}{c_s^2} -
    \frac{c_{\lambda}^m F_\lambda^m}{c_s^2} \delta^{kl}\Biggr) \Biggr]
    \quad ,
  \end{aligned}
\end{equation}
with $F_\lambda^i = -\Gamma^{i}_{jk} \xi_\lambda^j \xi_\lambda^k$ and
$\delta^{kl}$ is the Kronecker delta. In the presence of an external
force $F_{\text{ext}}$, this simply extends to $F_\lambda^i
\rightarrow F_\lambda^i + F^i_{\text{ext}\lambda}$.

In order to recover the correct macroscopic fluid equations, via a
Chapman-Enskog expansion, the other moments, Eq.~\eqref{moments}, also
need to be reproduced. A straightforward calculation shows that the
equilibrium distribution function $f_\lambda^{\rm eq}$ meets the
requirement. The shear viscosity of the fluid can also be calculated
as $\mu = \rho (\tau - 1/2) c_s^2 \delta t$. In this way one can
calculate the fluid motion in spaces with arbitrary local curvatures.

\section{Convergence Study}

\begin{figure}
  \centering
  \includegraphics[scale=0.35]{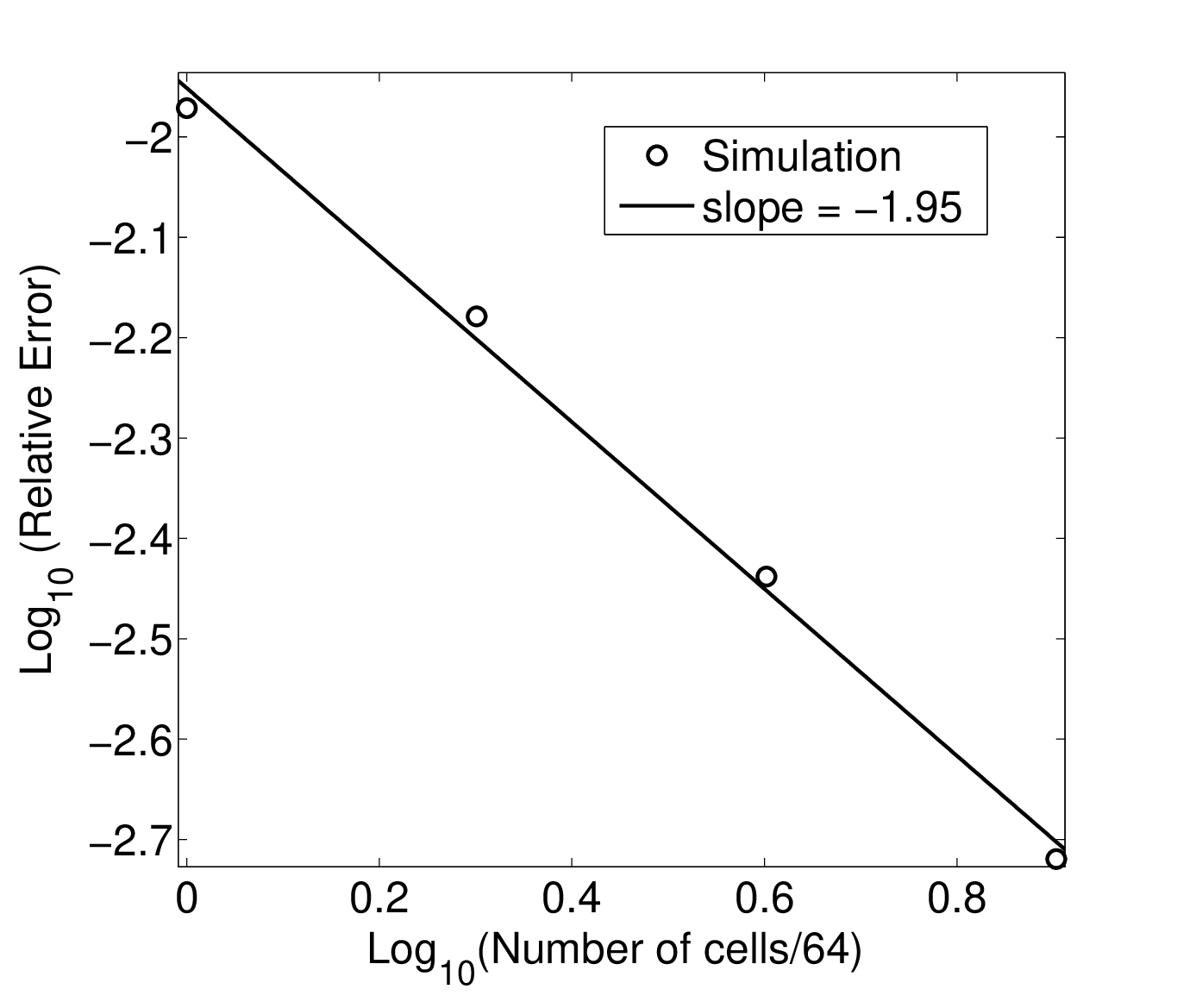}
  \caption{Relative convergence error as a function of the number of
    grid points. Here, the relative error is calculated by taken the
    mean value of the relative errors at every location grid point.
  }\label{convergenceerror}
\end{figure}
To check the convergence of the model, we simulate the Poiseuille
profile for the velocity on a two-dimensional ring. For this purpose,
we use the metric tensor in polar coordinates, $g_{rr} = 1$,
$g_{\theta \theta} = r^2$, and $g_{zz} = 1$, where $r$ is the radial
coordinate, $\theta$ is the azimuthal angle, and $z$ the axial
coordinate. Thus, the non-vanishing Christoffel symbols for this
metric are given by, $\Gamma^r_{\theta \theta} = -r$, and
$\Gamma^{\theta}_{r\theta} = \Gamma^{\theta}_{\theta r} = 1/r$.

Our system consists of a two-dimensional ring with inner radius $a$
and outer one $b$. On this ring, we impose a constant force $f_a$ in
the $\theta$-direction. For the simulation we choose $\tau = 0.6$. The
forcing term $f_a$ is set to $0.05$.  All numbers are expressed in
numerical units. The inner radius of the ring is taken as $a = 1.0$
and the outer radius as $b = 1.064$. We have taken periodic boundary
conditions in the direction $\theta$ and $z$, and free boundary
conditions at $r=1.0$ and $r=1.064$.

To obtain a quantitative measure of the convergence we use the
Richardson extrapolation method \cite{rich1,rich2}. In this method,
given any quantity $A(\delta x)$ that depends on a size step $\delta
x$, we can make an estimation of order $n$ of the exact solution $A$
by using
\begin{equation}\label{richardson1}
  A = \lim_{\delta x \rightarrow 0} A(\delta x) \approx \frac{2^n A\left (\frac{\delta x}{2}\right ) - A(\delta x)}{2^n - 1} + O(\delta x^{n+1})\quad ,
\end{equation}
with errors $O(\delta x^{n+1})$ of order $n+1$. Thus the relative
error between the value $A(\delta x)$ and the ``exact'' solution $A$
can be calculated by
\begin{equation}\label{richardson2}
  E_r(\delta x) = \left |\frac{A(\delta x) - A}{A} \right | \quad .
\end{equation}
In our case, the quantity $A$ is the fluid density $\rho$, when the
fluid reaches the steady state, and we set up $n=2$. Indeed, the
relative error with respect to the ``exact solution'' decreases
rapidly with increasing grid resolution (see
Fig.~\ref{convergenceerror}) and we can see that the present scheme
exhibits a near second-order convergence. This is basically in line
with the convergence properties of classical LB schemes.

\section{Validation}

To provide numerical validation of our model we study the
Taylor-Couette instability, which develops between two concentric
rotating cylinders. We calculate the critical Reynolds number, $Re_c$,
which characterizes the transition between stable Couette flow and
Taylor vortex flow. To this purpose, we use the metric tensor for
cylindrical coordinates $(r, \theta, z)$, $g_{rr} = 1$, $g_{\theta
  \theta} = r^2$, and $g_{zz} = 1$, where $r$ is the radial
coordinate, $\theta$ is the azimuthal angle, and $z$ the axial
coordinate. Thus, the non-vanishing Christoffel symbols for this
metric are given by $\Gamma^r_{\theta \theta} = -r$, and
$\Gamma^{\theta}_{r\theta} = \Gamma^{\theta}_{\theta r} = 1/r$.

In our system, the inner cylinder has radius $a$ and the outer one
radius $b$. We performed several simulations, by varying the Reynolds
number for different aspect ratios $\eta = a/b$. The Reynolds number,
assuming that the outer cylinder is fixed, can be defined as $Re = (a
\delta/\nu) d\theta/dt$ where $d\theta/dt$ is the angular speed of the
inner cylinder and $\delta = b - a$. The inner radius $a$ is always
set to $a=1$, and for a given value of $\eta$, the outer radius $b$
and $\delta$ are calculated. In order to vary $Re$, at fixed $\eta$,
we change the angular velocity of the inner cylinder. For this
simulation, we use a rectangular lattice of $128\times 1\times 256$
cells and choose $\tau = 1$ (all values are given in numerical units).
We use periodic boundary conditions in the $\theta$ and $z$
coordinates. At $r=a$ and $r=b$ boundaries, we have used free boundary
conditions, together with a condition to impose the respective angular
velocity at each boundary by evaluating the equilibrium function with
those values. Note that the boundary conditions can be implemented as
if they referred to a cartesian geometry, due to the use of
contravariant coordinates, leading to an approximation-free
representation of curved geometries. For smooth manifolds, the new
scheme is about three times slower than a standard cartesian version,
which is mainly due to the calculation of the metric and curvature
terms, as well as to the use of third order equilibria to enhance
stability.  Clearly, the advantage of the present scheme lies in the
treatment of complex manifolds which would require very high cartesian
grid resolution.

In Fig.~\ref{taylor:fig1}, we can observe the critical Reynolds number
as a function of $\eta$, as predicted by the simulation and compared
with the theoretical values from Ref.~\cite{transition}, finding
excellent agreement. We have implemented the same simulation using a
lattice size $64\times 1\times 256$ cells in order to study the
influence of the boundary conditions, and we found an error of around
$2.5\%$ respect to the theoretical values, which is a clear evidence
of the sensitivity to the boundary condition implementation. We have
been able to simulate Reynolds number of around $7000$ by using
$\tau=0.55$. Note that our model works in contravariant coordinates
and due to the presence of a metric tensor, the time step is not
necessarily unity.  For this reason, even when the relaxation time is
not small, the computed kinematic viscosity can achieve very small
values, leading to large Reynolds numbers.

For the case of two rotating spheres, we consider the inner sphere
with radius $a$ and the outer one with radius $b$. We use standard
spherical coordinates $(r,\phi,\theta)$, being $r$ the radial, $\phi$
the azimuthal, and $\theta$ the polar coordinates. The non-vanishing
components of the metric tensor are $g_{rr} = 1$, $g_{\phi \phi} = r^2
\sin^2(\theta)$, and $g_{\theta \theta} = r^2$.  The Christoffel
symbols can be calculated from the metric tensor by using standard
differential geometry relations. Note that our simulation region does
not include the poles because there, the determinant of the metric
tensor becomes zero and therefore it is not possible to calculate its
inverse. To circumvent this problem, we simulate the region $\theta \in (\pi/6,
5\pi/6)$. We set $\tau = 0.8$ and use a lattice of size $32\times 1
\times 384$. In order to vary the Reynolds number, we change the
azimuthal velocity $d\phi/dt$. The boundary conditions have been
chosen periodic for the case of $\phi$, and fixed for the case of $r$
and $\theta$. In the inset (left) of Fig.~\ref{taylor:fig1}, we show
the critical Reynolds number for different configurations which is in
good agreement with the experimental values given in Ref.
~\cite{tay3}. In this figure, we can also observe the radial and polar
components of the velocity, and see that there are two small vortices
located at the equator and two large ones at high and low latitudes,
in agreement with experiments and other numerical simulations
\cite{tay3,taylorsphe1,taylorsphe2}.

\begin{figure}
  \centering
  \includegraphics[scale=0.35]{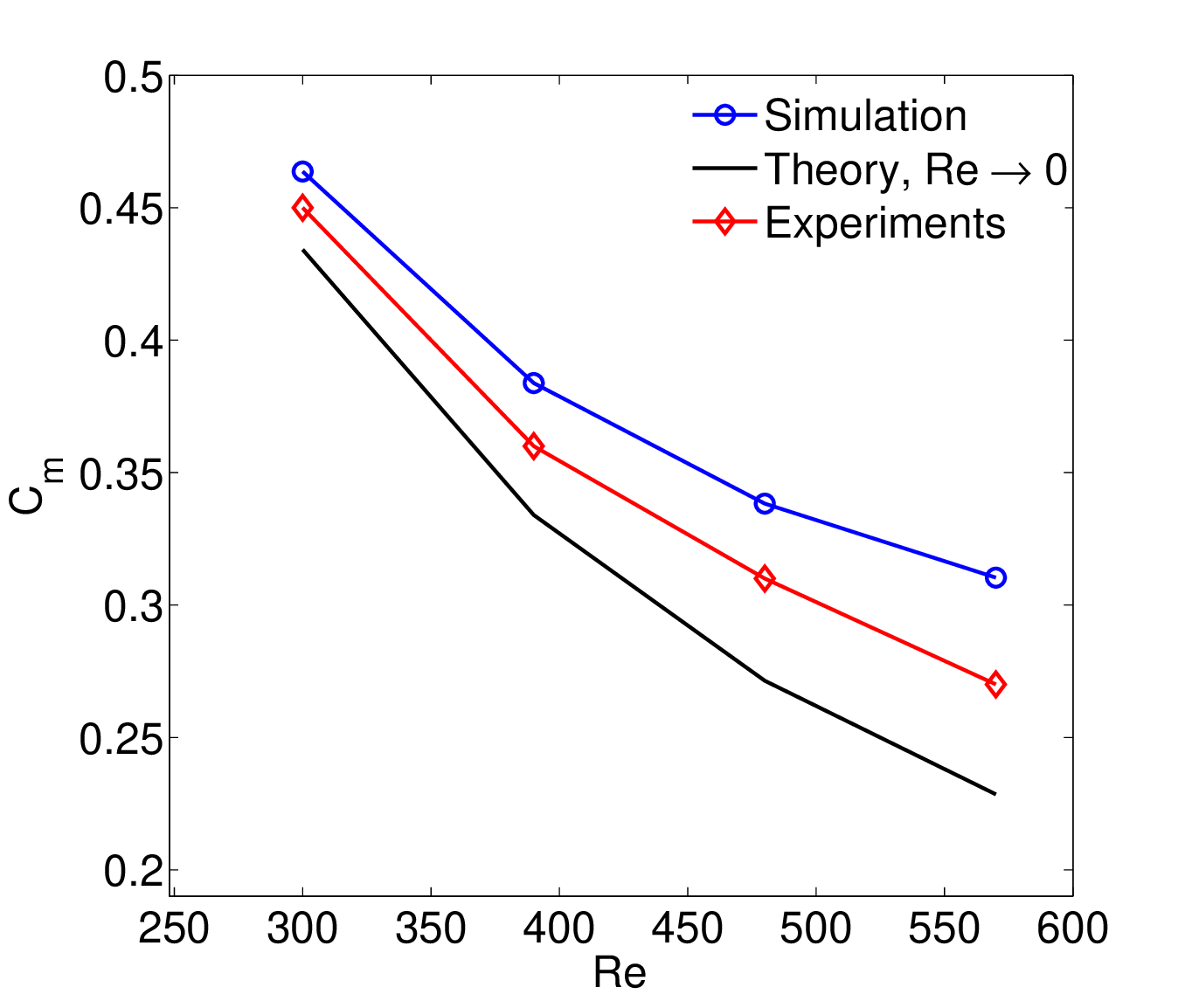}
  \caption{Torque coefficient as a function of the Reynolds number for
    the case of two concentric rotating spheres. The theoretical
    result has been taken from Ref.~\cite{torque1}, and the
    experimental data have been collected from Ref.~\cite{torque1,
      torque2}. }\label{torque}
\end{figure}
We have also measured the torque coefficient defined by,
\begin{equation}
  T_r = 2\pi a^3 \int_0^{\pi} \sigma_{r\phi} \sin^2(\theta) d\theta \quad ,
\end{equation}
where $\sigma_{r\phi}$ is the shear stress tensor, which in the
context of lattice kinetic theory can be calculated by,
\begin{equation}\label{sheartensor}
  \sigma^{\alpha \beta} = \left (1 - \frac{1}{2\tau} \right)\sum_{\lambda}^{41} (f_\lambda - f_\lambda^{\rm eq}) c_{\lambda}^{\alpha} c_{\lambda}^{\beta} \quad .
\end{equation}

The torque coefficient is then computed via the following relation
\cite{torque2}
\begin{equation}
C_m = \frac{T_r}{\frac{1}{2}\rho a^5 \left(\frac{d\phi}{dt}\right)^2 } \quad .
\end{equation}

In Fig.~\ref{torque}, we show the comparison between our results, the
theory for $Re \rightarrow 0$, and the experiments. We find good
agreement with the experiments. The small discrepancy can be due to
the approximation taken in Eq.~\eqref{sheartensor} and the
implementation of the boundary condition.

In order to study the Taylor-Couette instability for the case of two
concentric rotating tori, which to our knowledge has never been done
before, we use a lattice of size $64\times 128\times 64$ cells in the
orthogonal coordinate system of the torus, $(r,u,v)$, being $r$ the
radial, $u$ the axial, and $v$ the tangential coordinates. The
Christoffel symbols and the components of the metric tensor can be
readily calculated from differential geometry relations. The major
radius of the tori has been taken as $4.0$, in numerical units. The
other parameters are the same as in the previous simulations, and to
vary the Reynolds number we change the tangential velocity $dv/dt$. In
this case, $a$ and $b$ are the minor radii of the inner and outer
tori, respectively. We use periodic boundary conditions for the
coordinates $u$ and $v$, and fixed boundaries for $r$. In addition,
the critical Reynolds numbers for different configurations can be
observed in Fig.~\ref{taylor:fig1}, showing values around $10\%$
larger than for the case of cylinders.

\bibliography{report}

\begin{thebibliography}{28}%
\makeatletter
\providecommand \@ifxundefined [1]{%
 \@ifx{#1\undefined}
}%
\providecommand \@ifnum [1]{%
 \ifnum #1\expandafter \@firstoftwo
 \else \expandafter \@secondoftwo
 \fi
}%
\providecommand \@ifx [1]{%
 \ifx #1\expandafter \@firstoftwo
 \else \expandafter \@secondoftwo
 \fi
}%
\providecommand \natexlab [1]{#1}%
\providecommand \enquote  [1]{``#1''}%
\providecommand \bibnamefont  [1]{#1}%
\providecommand \bibfnamefont [1]{#1}%
\providecommand \citenamefont [1]{#1}%
\providecommand \href@noop [0]{\@secondoftwo}%
\providecommand \href [0]{\begingroup \@sanitize@url \@href}%
\providecommand \@href[1]{\@@startlink{#1}\@@href}%
\providecommand \@@href[1]{\endgroup#1\@@endlink}%
\providecommand \@sanitize@url [0]{\catcode `\\12\catcode `\$12\catcode
  `\&12\catcode `\#12\catcode `\^12\catcode `\_12\catcode `\%12\relax}%
\providecommand \@@startlink[1]{}%
\providecommand \@@endlink[0]{}%
\providecommand \url  [0]{\begingroup\@sanitize@url \@url }%
\providecommand \@url [1]{\endgroup\@href {#1}{\urlprefix }}%
\providecommand \urlprefix  [0]{URL }%
\providecommand \Eprint [0]{\href }%
\providecommand \doibase [0]{http://dx.doi.org/}%
\providecommand \selectlanguage [0]{\@gobble}%
\providecommand \bibinfo  [0]{\@secondoftwo}%
\providecommand \bibfield  [0]{\@secondoftwo}%
\providecommand \translation [1]{[#1]}%
\providecommand \BibitemOpen [0]{}%
\providecommand \bibitemStop [0]{}%
\providecommand \bibitemNoStop [0]{.\EOS\space}%
\providecommand \EOS [0]{\spacefactor3000\relax}%
\providecommand \BibitemShut  [1]{\csname bibitem#1\endcsname}%
\let\auto@bib@innerbib\@empty
\bibitem [{\citenamefont {Landau}\ and\ \citenamefont
  {Lifshitz}(1962)}]{landau}%
  \BibitemOpen
  \bibfield  {author} {\bibinfo {author} {\bibfnamefont {L.~D.}\ \bibnamefont
  {Landau}}\ and\ \bibinfo {author} {\bibfnamefont {E.~M.}\ \bibnamefont
  {Lifshitz}},\ }\href@noop {} {\emph {\bibinfo {title} {The classical theory
  of fields, by L. D. Landau and E. M. Lifshitz.}}},\ \bibinfo {edition} {rev.
  2d ed.}\ ed.\ (\bibinfo  {publisher} {Pergamon Press; Addison-Wesley Pub.
  Co., Oxford, Reading, Mass.,},\ \bibinfo {year} {1962})\ p.\ \bibinfo {pages}
  {404 p.}\BibitemShut {Stop}%
\bibitem [{\citenamefont {Seychelles}\ \emph {et~al.}(2008)\citenamefont
  {Seychelles}, \citenamefont {Amarouchene}, \citenamefont {Bessafi},\ and\
  \citenamefont {Kellay}}]{soap}%
  \BibitemOpen
  \bibfield  {author} {\bibinfo {author} {\bibfnamefont {F.}~\bibnamefont
  {Seychelles}}, \bibinfo {author} {\bibfnamefont {Y.}~\bibnamefont
  {Amarouchene}}, \bibinfo {author} {\bibfnamefont {M.}~\bibnamefont
  {Bessafi}}, \ and\ \bibinfo {author} {\bibfnamefont {H.}~\bibnamefont
  {Kellay}},\ }\href {\doibase 10.1103/PhysRevLett.100.144501} {\bibfield
  {journal} {\bibinfo  {journal} {Phys. Rev. Lett.}\ }\textbf {\bibinfo
  {volume} {100}},\ \bibinfo {pages} {144501} (\bibinfo {year}
  {2008})}\BibitemShut {NoStop}%
\bibitem [{\citenamefont {Priest}(1984)}]{solar}%
  \BibitemOpen
  \bibfield  {author} {\bibinfo {author} {\bibfnamefont {E.}~\bibnamefont
  {Priest}},\ }\href {http://books.google.ch/books?id=sZylLdvm7lMC} {\emph
  {\bibinfo {title} {Solar magneto-hydrodynamics}}},\ Geophysics and
  astrophysics monographs\ (\bibinfo  {publisher} {D. Reidel Pub. Co.},\
  \bibinfo {year} {1984})\BibitemShut {NoStop}%
\bibitem [{\citenamefont {Mullin}\ and\ \citenamefont {Blohm}(2001)}]{tay1}%
  \BibitemOpen
  \bibfield  {author} {\bibinfo {author} {\bibfnamefont {T.}~\bibnamefont
  {Mullin}}\ and\ \bibinfo {author} {\bibfnamefont {C.}~\bibnamefont {Blohm}},\
  }\href {\doibase DOI:10.1063/1.1329906} {\bibfield  {journal} {\bibinfo
  {journal} {Phys. of Fluids}\ }\textbf {\bibinfo {volume} {13}},\ \bibinfo
  {pages} {136} (\bibinfo {year} {2001})}\BibitemShut {NoStop}%
\bibitem [{\citenamefont {Di~Prima}\ and\ \citenamefont
  {Swinney}(1985)}]{transition}%
  \BibitemOpen
  \bibfield  {author} {\bibinfo {author} {\bibfnamefont {R.}~\bibnamefont
  {Di~Prima}}\ and\ \bibinfo {author} {\bibfnamefont {H.}~\bibnamefont
  {Swinney}},\ }in\ \href@noop {} {\emph {\bibinfo {booktitle} {Hydrodynamic
  Instabilities and the Transition to Turbulence}}},\ \bibinfo {series} {Topics
  in Applied Physics}, Vol.~\bibinfo {volume} {45},\ \bibinfo {editor} {edited
  by\ \bibinfo {editor} {\bibfnamefont {H.}~\bibnamefont {Swinney}}\ and\
  \bibinfo {editor} {\bibfnamefont {J.}~\bibnamefont {Gollub}}}\ (\bibinfo
  {publisher} {Springer Berlin / Heidelberg},\ \bibinfo {year} {1985})\ pp.\
  \bibinfo {pages} {139--180}\BibitemShut {NoStop}%
\bibitem [{\citenamefont {Bartels}(1982{\natexlab{a}})}]{taylorsphe1}%
  \BibitemOpen
  \bibfield  {author} {\bibinfo {author} {\bibfnamefont {F.}~\bibnamefont
  {Bartels}},\ }\href@noop {} {\bibfield  {journal} {\bibinfo  {journal} {J.
  Fluid Mech.}\ }\textbf {\bibinfo {volume} {119}},\ \bibinfo {pages} {1 }
  (\bibinfo {year} {1982}{\natexlab{a}})}\BibitemShut {NoStop}%
\bibitem [{\citenamefont {Schrauf}(1986)}]{tay3}%
  \BibitemOpen
  \bibfield  {author} {\bibinfo {author} {\bibfnamefont {G.}~\bibnamefont
  {Schrauf}},\ }\href {\doibase 10.1017/S0022112086000150} {\bibfield
  {journal} {\bibinfo  {journal} {Journal of Fluid Mechanics}\ }\textbf
  {\bibinfo {volume} {166}},\ \bibinfo {pages} {287} (\bibinfo {year}
  {1986})}\BibitemShut {NoStop}%
\bibitem [{\citenamefont {Hirsch}\ and\ \citenamefont {Hirsch}(2007)}]{CFD1}%
  \BibitemOpen
  \bibfield  {author} {\bibinfo {author} {\bibfnamefont {C.}~\bibnamefont
  {Hirsch}}\ and\ \bibinfo {author} {\bibfnamefont {C.}~\bibnamefont
  {Hirsch}},\ }\href@noop {} {\emph {\bibinfo {title} {Numerical Computation of
  Internal and External Flows: Fundamentals of Computational Fluid
  Dynamics}}},\ \bibinfo {series} {Butterworth Heinemann}\ No.\ \bibinfo
  {number} {Bd. 1}\ (\bibinfo  {publisher} {Butterworth-Heinemann},\ \bibinfo
  {year} {2007})\BibitemShut {NoStop}%
\bibitem [{\citenamefont {Chen}\ \emph {et~al.}(2003)\citenamefont {Chen},
  \citenamefont {Kandasamy}, \citenamefont {Orszag}, \citenamefont {Shock},
  \citenamefont {Succi},\ and\ \citenamefont {Yakhot}}]{CFD2}%
  \BibitemOpen
  \bibfield  {author} {\bibinfo {author} {\bibfnamefont {H.}~\bibnamefont
  {Chen}}, \bibinfo {author} {\bibfnamefont {S.}~\bibnamefont {Kandasamy}},
  \bibinfo {author} {\bibfnamefont {S.}~\bibnamefont {Orszag}}, \bibinfo
  {author} {\bibfnamefont {R.}~\bibnamefont {Shock}}, \bibinfo {author}
  {\bibfnamefont {S.}~\bibnamefont {Succi}}, \ and\ \bibinfo {author}
  {\bibfnamefont {V.}~\bibnamefont {Yakhot}},\ }\href {\doibase
  10.1126/science.1085048} {\bibfield  {journal} {\bibinfo  {journal}
  {Science}\ }\textbf {\bibinfo {volume} {301}},\ \bibinfo {pages} {633}
  (\bibinfo {year} {2003})}\BibitemShut {NoStop}%
\bibitem [{\citenamefont {Marcus}\ and\ \citenamefont
  {Tuckerman}(1987)}]{taylorsphe2}%
  \BibitemOpen
  \bibfield  {author} {\bibinfo {author} {\bibfnamefont {P.}~\bibnamefont
  {Marcus}}\ and\ \bibinfo {author} {\bibfnamefont {L.}~\bibnamefont
  {Tuckerman}},\ }\href@noop {} {\bibfield  {journal} {\bibinfo  {journal} {J.
  Fluid Mech.}\ }\textbf {\bibinfo {volume} {185}} (\bibinfo {year}
  {1987})}\BibitemShut {NoStop}%
\bibitem [{\citenamefont {Baumgarte}\ and\ \citenamefont
  {Shapiro}(1998)}]{numRel1}%
  \BibitemOpen
  \bibfield  {author} {\bibinfo {author} {\bibfnamefont {T.~W.}\ \bibnamefont
  {Baumgarte}}\ and\ \bibinfo {author} {\bibfnamefont {S.~L.}\ \bibnamefont
  {Shapiro}},\ }\href@noop {} {\bibfield  {journal} {\bibinfo  {journal} {Phys.
  Rev. D}\ }\textbf {\bibinfo {volume} {59}},\ \bibinfo {pages} {024007}
  (\bibinfo {year} {1998})}\BibitemShut {NoStop}%
\bibitem [{\citenamefont {Schnetter}\ \emph {et~al.}(2004)\citenamefont
  {Schnetter}, \citenamefont {Hawley},\ and\ \citenamefont {Hawke}}]{numRel2}%
  \BibitemOpen
  \bibfield  {author} {\bibinfo {author} {\bibfnamefont {E.}~\bibnamefont
  {Schnetter}}, \bibinfo {author} {\bibfnamefont {S.~H.}\ \bibnamefont
  {Hawley}}, \ and\ \bibinfo {author} {\bibfnamefont {I.}~\bibnamefont
  {Hawke}},\ }\href@noop {} {\bibfield  {journal} {\bibinfo  {journal}
  {Classical and Quantum Gravity}\ }\textbf {\bibinfo {volume} {21}},\ \bibinfo
  {pages} {1465} (\bibinfo {year} {2004})}\BibitemShut {NoStop}%
\bibitem [{sup()}]{supp}%
  \BibitemOpen
  \href@noop {} {}\bibinfo {note} {See Supplemental Material at.}\BibitemShut
  {Stop}%
\bibitem [{\citenamefont {Love}\ and\ \citenamefont {Cianci}(2011)}]{donato}%
  \BibitemOpen
  \bibfield  {author} {\bibinfo {author} {\bibfnamefont {P.~J.}\ \bibnamefont
  {Love}}\ and\ \bibinfo {author} {\bibfnamefont {D.}~\bibnamefont {Cianci}},\
  }\href {\doibase 10.1098/rsta.2011.0097} {\bibfield  {journal} {\bibinfo
  {journal} {Philosophical Transactions of the Royal Society A: Mathematical,
  Physical and Engineering Sciences}\ }\textbf {\bibinfo {volume} {369}},\
  \bibinfo {pages} {2362} (\bibinfo {year} {2011})}\BibitemShut {NoStop}%
\bibitem [{\citenamefont {Sinitsyn}\ \emph {et~al.}(2011)\citenamefont
  {Sinitsyn}, \citenamefont {Dulov},\ and\ \citenamefont
  {Vedenyapin}}]{KineticBoltzmann}%
  \BibitemOpen
  \bibfield  {author} {\bibinfo {author} {\bibfnamefont {A.}~\bibnamefont
  {Sinitsyn}}, \bibinfo {author} {\bibfnamefont {E.}~\bibnamefont {Dulov}}, \
  and\ \bibinfo {author} {\bibfnamefont {V.}~\bibnamefont {Vedenyapin}},\
  }\href@noop {} {\emph {\bibinfo {title} {Kinetic Boltzmann, Vlasov and
  Related Equations}}}\ (\bibinfo  {publisher} {Elsevier},\ \bibinfo {year}
  {2011})\BibitemShut {NoStop}%
\bibitem [{\citenamefont {Wimmer}(1976)}]{torque1}%
  \BibitemOpen
  \bibfield  {author} {\bibinfo {author} {\bibfnamefont {M.}~\bibnamefont
  {Wimmer}},\ }\href@noop {} {\bibfield  {journal} {\bibinfo  {journal} {J.
  Fluid Mech.}\ }\textbf {\bibinfo {volume} {78}},\ \bibinfo {pages} {317}
  (\bibinfo {year} {1976})}\BibitemShut {NoStop}%
\bibitem [{\citenamefont {Bartels}(1982{\natexlab{b}})}]{torque2}%
  \BibitemOpen
  \bibfield  {author} {\bibinfo {author} {\bibfnamefont {F.}~\bibnamefont
  {Bartels}},\ }\href@noop {} {\bibfield  {journal} {\bibinfo  {journal} {J.
  Fluid Mech.}\ }\textbf {\bibinfo {volume} {119}},\ \bibinfo {pages} {1}
  (\bibinfo {year} {1982}{\natexlab{b}})}\BibitemShut {NoStop}%
\bibitem [{\citenamefont {Bini}\ \emph {et~al.}(2009)\citenamefont {Bini},
  \citenamefont {Jantzen},\ and\ \citenamefont {Stella}}]{BINI1}%
  \BibitemOpen
  \bibfield  {author} {\bibinfo {author} {\bibfnamefont {D.}~\bibnamefont
  {Bini}}, \bibinfo {author} {\bibfnamefont {R.~T.}\ \bibnamefont {Jantzen}}, \
  and\ \bibinfo {author} {\bibfnamefont {L.}~\bibnamefont {Stella}},\
  }\href@noop {} {\bibfield  {journal} {\bibinfo  {journal} {Classical and
  Quantum Gravity}\ }\textbf {\bibinfo {volume} {26}},\ \bibinfo {pages}
  {055009} (\bibinfo {year} {2009})}\BibitemShut {NoStop}%
\bibitem [{\citenamefont {Bini}\ \emph {et~al.}(2012)\citenamefont {Bini},
  \citenamefont {Gregoris},\ and\ \citenamefont {Succi}}]{BINI2}%
  \BibitemOpen
  \bibfield  {author} {\bibinfo {author} {\bibfnamefont {D.}~\bibnamefont
  {Bini}}, \bibinfo {author} {\bibfnamefont {D.}~\bibnamefont {Gregoris}}, \
  and\ \bibinfo {author} {\bibfnamefont {S.}~\bibnamefont {Succi}},\
  }\href@noop {} {\bibfield  {journal} {\bibinfo  {journal} {EPL (Europhysics
  Letters)}\ }\textbf {\bibinfo {volume} {97}},\ \bibinfo {pages} {40007}
  (\bibinfo {year} {2012})}\BibitemShut {NoStop}%
\bibitem [{\citenamefont {Mendoza}\ \emph {et~al.}(2010)\citenamefont
  {Mendoza}, \citenamefont {Boghosian}, \citenamefont {Herrmann},\ and\
  \citenamefont {Succi}}]{rlbPRL}%
  \BibitemOpen
  \bibfield  {author} {\bibinfo {author} {\bibfnamefont {M.}~\bibnamefont
  {Mendoza}}, \bibinfo {author} {\bibfnamefont {B.~M.}\ \bibnamefont
  {Boghosian}}, \bibinfo {author} {\bibfnamefont {H.~J.}\ \bibnamefont
  {Herrmann}}, \ and\ \bibinfo {author} {\bibfnamefont {S.}~\bibnamefont
  {Succi}},\ }\href@noop {} {\bibfield  {journal} {\bibinfo  {journal} {Phys.
  Rev. Lett.}\ }\textbf {\bibinfo {volume} {105}},\ \bibinfo {pages} {014502}
  (\bibinfo {year} {2010})}\BibitemShut {NoStop}%
\bibitem [{\citenamefont {Mendoza}\ \emph {et~al.}(2011)\citenamefont
  {Mendoza}, \citenamefont {Herrmann},\ and\ \citenamefont {Succi}}]{turbPRL}%
  \BibitemOpen
  \bibfield  {author} {\bibinfo {author} {\bibfnamefont {M.}~\bibnamefont
  {Mendoza}}, \bibinfo {author} {\bibfnamefont {H.~J.}\ \bibnamefont
  {Herrmann}}, \ and\ \bibinfo {author} {\bibfnamefont {S.}~\bibnamefont
  {Succi}},\ }\href@noop {} {\bibfield  {journal} {\bibinfo  {journal} {Phys.
  Rev. Lett.}\ }\textbf {\bibinfo {volume} {106}},\ \bibinfo {pages} {156601}
  (\bibinfo {year} {2011})}\BibitemShut {NoStop}%
\bibitem [{\citenamefont {Grad}(1949)}]{grad}%
  \BibitemOpen
  \bibfield  {author} {\bibinfo {author} {\bibfnamefont {H.}~\bibnamefont
  {Grad}},\ }\href {\doibase 10.1002/cpa.3160020402} {\bibfield  {journal}
  {\bibinfo  {journal} {Communications on Pure and Applied Mathematics}\
  }\textbf {\bibinfo {volume} {2}},\ \bibinfo {pages} {325} (\bibinfo {year}
  {1949})}\BibitemShut {NoStop}%
\bibitem [{\citenamefont {Martys}\ \emph {et~al.}(1998)\citenamefont {Martys},
  \citenamefont {Shan},\ and\ \citenamefont {Chen}}]{discLB1}%
  \BibitemOpen
  \bibfield  {author} {\bibinfo {author} {\bibfnamefont {N.~S.}\ \bibnamefont
  {Martys}}, \bibinfo {author} {\bibfnamefont {X.}~\bibnamefont {Shan}}, \ and\
  \bibinfo {author} {\bibfnamefont {H.}~\bibnamefont {Chen}},\ }\href {\doibase
  10.1103/PhysRevE.58.6855} {\bibfield  {journal} {\bibinfo  {journal} {Phys.
  Rev. E}\ }\textbf {\bibinfo {volume} {58}},\ \bibinfo {pages} {6855}
  (\bibinfo {year} {1998})}\BibitemShut {NoStop}%
\bibitem [{\citenamefont {Shan}\ and\ \citenamefont {He}(1998)}]{discLB2}%
  \BibitemOpen
  \bibfield  {author} {\bibinfo {author} {\bibfnamefont {X.}~\bibnamefont
  {Shan}}\ and\ \bibinfo {author} {\bibfnamefont {X.}~\bibnamefont {He}},\
  }\href {\doibase 10.1103/PhysRevLett.80.65} {\bibfield  {journal} {\bibinfo
  {journal} {Phys. Rev. Lett.}\ }\textbf {\bibinfo {volume} {80}},\ \bibinfo
  {pages} {65} (\bibinfo {year} {1998})}\BibitemShut {NoStop}%
\bibitem [{\citenamefont {Chikatamarla}\ and\ \citenamefont
  {Karlin}(2009)}]{karli}%
  \BibitemOpen
  \bibfield  {author} {\bibinfo {author} {\bibfnamefont {S.~S.}\ \bibnamefont
  {Chikatamarla}}\ and\ \bibinfo {author} {\bibfnamefont {I.~V.}\ \bibnamefont
  {Karlin}},\ }\href {\doibase 10.1103/PhysRevE.79.046701} {\bibfield
  {journal} {\bibinfo  {journal} {Phys. Rev. E}\ }\textbf {\bibinfo {volume}
  {79}},\ \bibinfo {pages} {046701} (\bibinfo {year} {2009})}\BibitemShut
  {NoStop}%
\bibitem [{\citenamefont {Karlin}\ \emph {et~al.}(1998)\citenamefont {Karlin},
  \citenamefont {Gorban}, \citenamefont {Succi},\ and\ \citenamefont
  {Boffi}}]{ELB0}%
  \BibitemOpen
  \bibfield  {author} {\bibinfo {author} {\bibfnamefont {I.~V.}\ \bibnamefont
  {Karlin}}, \bibinfo {author} {\bibfnamefont {A.~N.}\ \bibnamefont {Gorban}},
  \bibinfo {author} {\bibfnamefont {S.}~\bibnamefont {Succi}}, \ and\ \bibinfo
  {author} {\bibfnamefont {V.}~\bibnamefont {Boffi}},\ }\href {\doibase
  10.1103/PhysRevLett.81.6} {\bibfield  {journal} {\bibinfo  {journal} {Phys.
  Rev. Lett.}\ }\textbf {\bibinfo {volume} {81}},\ \bibinfo {pages} {6}
  (\bibinfo {year} {1998})}\BibitemShut {NoStop}%
\bibitem [{\citenamefont {Richardson}(1911)}]{rich1}%
  \BibitemOpen
  \bibfield  {author} {\bibinfo {author} {\bibfnamefont {L.~F.}\ \bibnamefont
  {Richardson}},\ }\href {\doibase 10.1098/rsta.1911.0009} {\bibfield
  {journal} {\bibinfo  {journal} {Philosophical Transactions of the Royal
  Society of London. Series A, Containing Papers of a Mathematical or Physical
  Character}\ }\textbf {\bibinfo {volume} {210}},\ \bibinfo {pages} {307}
  (\bibinfo {year} {1911})}\BibitemShut {NoStop}%
\bibitem [{\citenamefont {Richardson}\ and\ \citenamefont
  {Gaunt}(1927)}]{rich2}%
  \BibitemOpen
  \bibfield  {author} {\bibinfo {author} {\bibfnamefont {L.~F.}\ \bibnamefont
  {Richardson}}\ and\ \bibinfo {author} {\bibfnamefont {J.~A.}\ \bibnamefont
  {Gaunt}},\ }\href {\doibase 10.1098/rsta.1927.0008} {\bibfield  {journal}
  {\bibinfo  {journal} {Philosophical Transactions of the Royal Society of
  London. Series A, Containing Papers of a Mathematical or Physical Character}\
  }\textbf {\bibinfo {volume} {226}},\ \bibinfo {pages} {299} (\bibinfo {year}
  {1927})}\BibitemShut {NoStop}%
\end{thebibliography}%

\end{document}